\documentclass[aip,amsmath,amssymb,twocolumn,]{revtex4}                

\usepackage{textcase}
\usepackage{graphicx}
\usepackage{dcolumn}
\usepackage{bm}
\usepackage{amsmath,amsthm,amssymb,mathrsfs}

\begin{document}

\title{Out-of-plane magnetization oscillation in spin Hall device assisted by field-like torque}

\author{Tomohiro Taniguchi}
\email{tomohiro-taniguchi@aist.go.jp}

\affiliation{ 
National Institute of Advanced Industrial Science and Technology (AIST), Research Center for Emerging Computing Technologies, Tsukuba 305-8568, Japan}

\date{\today}%

\begin{abstract}
An excitation of a large-amplitude out-of-plane magnetization oscillation in a ferromagnet by the spin Hall effect is of great interest 
for practical applications such as microwave generator and neuromorphic computing. 
However, both experimental and theoretical works have revealed that only small-amplitude oscillation around an in-plane easy axis can be excited via the spin Hall effect. 
Here, we propose that an out-of-plane oscillation can be excited due to an assistance of field-like torque. 
We focus on an in-plane magnetized ferromagnet with an easy axis parallel to current direction. 
We notice that the field-like torque with an appropriate sign provides an additional field modifying the dynamic trajectory of the magnetization and drives the auto-oscillation. 
The condition on the sign of the field-like torque is satisfied for typical nonmagnet used in spin Hall devices such as tungsten. 
\end{abstract}

\maketitle


Spin-orbit interaction in nonmagnet scatters spin-up and spin-down electrons to the opposite direction 
and generates pure spin current, which is known as the spin Hall effect \cite{dyakonov71,hirsch99,kato04,kimura07,ando08}. 
The pure spin current injected into an adjacent ferromagnet excites spin-transfer torque on its magnetization and induces magnetization dynamics. 
Whereas the magnetization switching between two stable states has been mainly studied for the purpose of three-terminal non-volatile memory 
\cite{liu12a,lee13,cubukcu14,yu14,garello14,you15,torrejon15,lau16,brink16,oh16}, 
another fascinating dynamics induced by the spin-transfer torque is an auto-oscillation of the magnetization. 
It was demonstrated experimentally that the spin Hall effect can induces the oscillation in in-plane magnetized ferromagnet \cite{liu12b}. 
The result indicates a possibility to apply the spin-Hall devices to other practical devices such as microwave generators and neuromorphic devices \cite{demidov14,awad17,kudo17}. 


An intriguing challenge in the study of spin-transfer torque induced auto-oscillation is an excitation of a large-amplitude out-of-plane oscillation 
because the emitted power outputted from the oscillator is proportional to the oscillation amplitude. 
Unfortunately, however, the oscillation amplitude of the auto-oscillation by the spin Hall effect was small because the oscillation occurred around an in-plane easy axis \cite{liu12b}. 
Although a large amplitude oscillation might be possibly excited in the in-plane magnetized ferromagnet, an external magnetic field  will be necessary \cite{kiselev03},  
which is undesirable for practical purpose. 
Another proposal requires time-dependent input and accurate control of the magnetic anisotropy energy, 
where both minima and maxima of the energy landscape locate in the in-plane direction whereas saddle points locate out-of-plane direction \cite{shirokura20}. 
It was also proved theoretically that the spin Hall effect does not excite an out-of-plane oscillation in a perpendicularly magnetized ferromagnet \cite{taniguchi15,taniguchi15PRB}. 
Regarding such a current status, it is of great interest to propose a method to excite an out-of-plane magnetization oscillation by the spin Hall effect. 


The purpose of this work is to propose a geometry and conditions to excite an out-of-plane magnetization oscillation by the spin Hall effect theoretically. 
We focus on an in-plane magnetized ferromagnet, as in the case of previous work \cite{liu12b}, however, the geometry is different. 
The previous work used a ferromagnet having an easy axis orthogonal to current direction. 
According to Ref. \cite{fukami16}, we call this geometry type Y, for convention. 
On the other hand, we focus on a ferromagnet with its easy axis parallel to the current, which was named as type X \cite{fukami16}. 
In addition, the field-like torque is taken into account in equation of motion of the magnetization. 
The field-like torque provides an effective magnetic field pointing in a direction of the in-plane hard axis and modifies an energy landscape. 
As a result of the competition between the spin-transfer torque and the damping torque including the contribution from the field-like torque, 
an out-of-plane magnetization oscillation can be excited when the field-like torque has an appropriate sign. 
The condition on the sign of the field-like torque is satisfied in, for example, tungsten \cite{lau17}. 
Therefore, it will be possible to examine the present proposal experimentally. 




\begin{figure}
\centerline{\includegraphics[width=1.0\columnwidth]{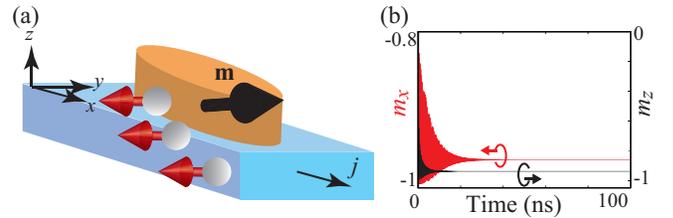}}
\caption{
        (a) Schematic illustration of the system. 
           Electric current density $j$ flowing in the $x$ direction in the bottom nonmagnet generates pure spin current with spin polarization (red arrow) in the $y$ direction, 
           which is injected into the top ferromagnet and excites spin-transfer torque on the magnetization $\mathbf{m}$. 
        (b) Time evolutions of $m_{x}$ (red) for the current density of $j=90$ MA/cm${}^{2}$ and $m_{z}$ (black) for $j=340$ MA/cm${}^{2}$. 
         \vspace{-3ex}}
\label{fig:fig1}
\end{figure}



Figure \ref{fig:fig1}(a) is a schematic illustration of the system under consideration. 
The $x$ axis is parallel to current direction, whereas the $z$ axis points to the perpendicular direction. 
The current flowing in the bottom nonmagnet generates pure spin current injected into the top ferromagnet and excites spin-transfer torque on magnetization. 
We use macrospin assumption to describe the magnetization dynamics excited by the spin-transfer torque because its validity was experimentally confirmed \cite{fukami16}. 
The Landau-Lifshitz-Gilbert (LLG) equation of the magnetization is given by 
\begin{equation}
\begin{split}
  \frac{d \mathbf{m}}{dt}
  =&
  -\gamma
  \mathbf{m}
  \times
  \mathbf{H}
  +
  \alpha
  \mathbf{m}
  \times
  \frac{d \mathbf{m}}{dt} 
\\
  &-
  \frac{\gamma \hbar \vartheta j}{2eMd}
  \mathbf{m}
  \times
  \left(
    \mathbf{e}_{y}
    \times
    \mathbf{m}
  \right)
  -
  \frac{\gamma \hbar \beta \vartheta j}{2eMd}
  \mathbf{m}
  \times
  \mathbf{e}_{y}, 
  \label{eq:LLG}
\end{split}
\end{equation}
where $\mathbf{m}$ and $\mathbf{e}_{y}$ are the unit vectors pointing in the magnetization and $y$ directions, respectively. 
The gyromagnetic ratio and the Gilbert damping constant are denoted as $\gamma$ and $\alpha$, respectively. 
The magnetic field consists of shape and interfacial crystalline anisotropy fields as 
\begin{equation}
  \mathbf{H}
  =
  \begin{pmatrix}
    -4\pi M N_{x} m_{x} \\
    -4\pi M N_{y} m_{y} \\
    (H_{\rm K}-4\pi M N_{z}) m_{z}
  \end{pmatrix}, 
\end{equation}
where $N_{\ell}$ ($\ell=x,y,z$) is the demagnetization coefficient, 
whereas $H_{\rm K}$ represents the interfacial contribution to the perpendicular magnetic anisotropy \cite{yakata09,ikeda10,kubota12}. 
Since we are interested in a ferromagnet having its easy axis along the $x$ direction, 
$N_{x}<N_{y}<N_{z}$ and $H_{\rm K}-4\pi MN_{z}<0$. 
The saturation magnetization and thickness of the ferromagnet are denoted as $M$ and $d$, respectively. 
The spin Hall angle, which corresponds to the charge-to-spin conversion efficiency, is $\vartheta$, whereas the dimensionless parameter $\beta$ is the ratio of the field-like torque to the damping-like torque, 
which correspond to the fourth and third terms on the right-hand side of Eq. (\ref{eq:LLG}). 
The electric current density flowing in the bottom nonmagnet is $j$. 
The values of the parameters are derived from recent experiment \cite{shiokawa21} as 
$M=1540$ emu/cm${}^{3}$, $H_{\rm K}=14.4$ kOe, $\gamma=1.764 \times 10^{7}$ rad/(Oe s), $\alpha=0.04$, and $d=2$ nm. 
The long and short diameters of the ferromagnet are 297 nm in the $x$ direction and 122 nm in the $y$ direction. 
Then, according to Ref. \cite{beleggia05}, the demagnetization coefficients are estimated to be $N_{x}=0.008925$, $N_{y}=0.031527$, and $N_{z}=0.959548$. 
The spin Hall angle is $\vartheta=-0.338$ for tungsten \cite{shiokawa21}. 
The initial state is set to be an stable state, $\mathbf{m}=-\mathbf{e}_{x}$. 
The fourth-order Runge-Kutta method is applied to solve Eq. (\ref{eq:LLG}) numerically. 
Whereas we focus on the dynamics at zero temperature in the following, 
the following conclusions are unchanged even at finite temperature, as shown in supplementary material. 


As discussed below, the parameter $\beta$ of the field-like torque plays a key role to excite an auto-oscillation. 
The parameter $\beta$ includes not only the contributions from bulk spin Hall effect and interfacial spin-orbit interaction but also the contribution from the Oersted field. 
Although several efforts have been made to separate these contributions by using a variety of experimental techniques \cite{lau17,pai12,kim13,nakayama13,althammer13,hayashi14,kim16,takeuchi18}, 
the magnetization dynamics are affected by all of them. 
Therefore, we regard $\beta$ in this study as the total value including these contributions. 
Accordingly, the sign of $\beta$ depends on materials through the bulk and interfacial contributions, as well as structures. 
For example, $\beta$ of systems with platinum, tungsten, and tantalum are positive \cite{lau17}. 
In addition, the magnitude of $\beta$ could be on the order of $1$ \cite{kim13}. 
Therefore, we use $\beta=0.3$ in the following, whereas the results for different values of $\beta$ are shown in supplementary material. 



Figures \ref{fig:fig1}(b) shows the time evolutions of $m_{x}$ (red) for the current density of $j=90$ MA/cm${}^{2}$ 
and $m_{z}$ (black) for $j=340$ MA/cm${}^{2}$. 
The magnetization slightly deviates from the initial state but stays close to the easy axis when the current density is small. 
On the other hand, the magnetization moves to the $z$ direction when the current is large. 
In both cases, the magnetization does not show oscillation. 
In the middle current region, however, we find auto-oscillation of the magnetization as follows. 

\begin{figure*}
\centerline{\includegraphics[width=2.0\columnwidth]{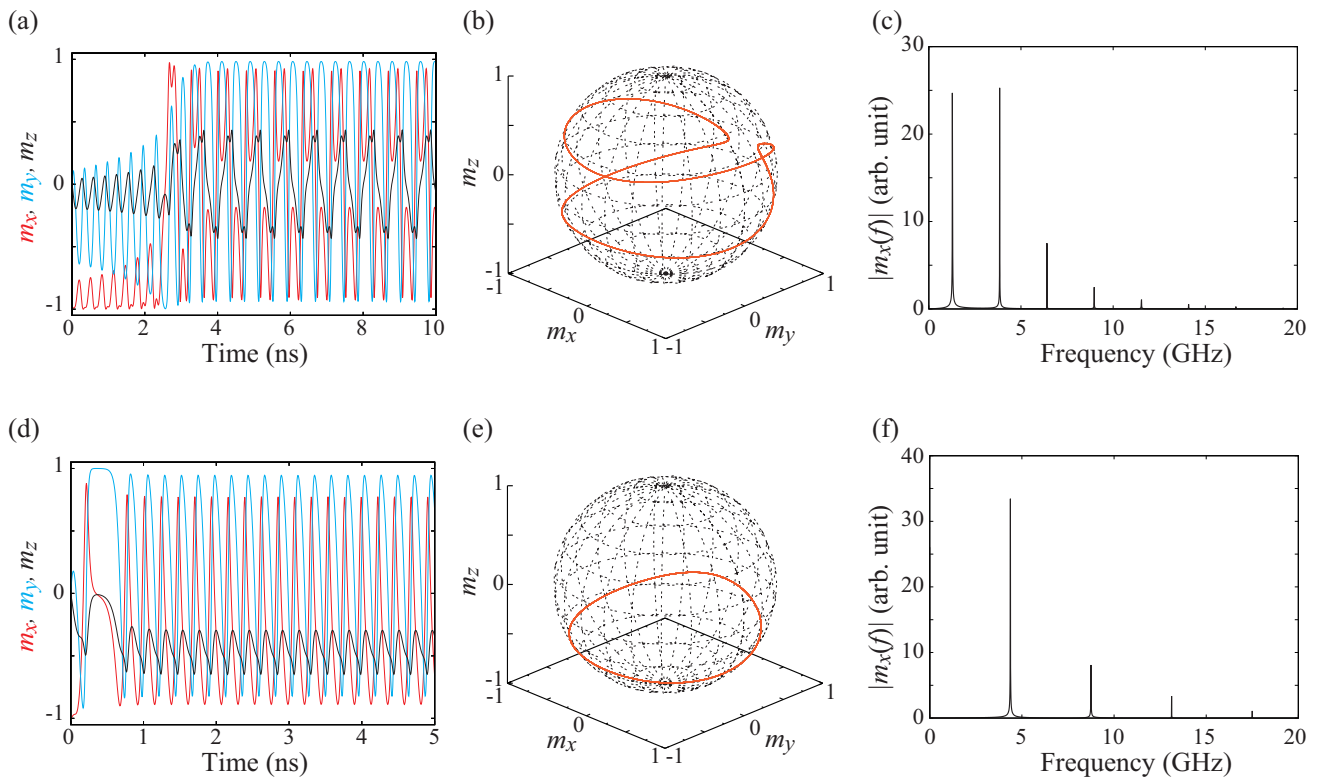}}
\caption{
         (a) Time evolutions of $m_{x}$ (red), $m_{y}$ (blue), and $m_{z}$ (black), 
         (b) the dynamical trajectory in a steady state, and 
         (c) the Fourier transformation of $m_{x}$ for the current density of $j=97$ MA/cm${}^{2}$. 
         Similar plots with different current density, $j=160$ MA/cm${}^{2}$, are shown in (d)-(f). 
         \vspace{-3ex}}
\label{fig:fig2}
\end{figure*}

Figure \ref{fig:fig2}(a) shows the time evolutions of $m_{x}$ (red), $m_{y}$ (blue), and $m_{z}$ (black) for the current density of $j=97$ MA/cm${}^{2}$. 
The magnetization finally saturates to an oscillation state after deviating from the initial state. 
Figure \ref{fig:fig2}(b) shows the dynamical trajectory of the magnetization in a steady state. 
It shows the magnetization oscillates around the negative $y$ axis. 
Figure \ref{fig:fig2}(c) shows the Fourier transformation $|m_{x}(f)|$ of $m_{x}$. 
The spectrum has multiple peaks reflecting a complex trajectory shown in Fig. \ref{fig:fig2}(b). 
The first main peak appears at $1.3$ GHz, which is higher than that found in two-terminal spintronics devices without external magnetic field \cite{zeng12}. 
Note that the oscillation occurs around an effective field associated by the field-like torque. 
In fact, by taking into the fact that $\beta\vartheta<0$, 
the field-like torque in Eq. (\ref{eq:LLG}) can be regarded as a torque due to magnetic field in the negative $y$ direction. 
Therefore, we can conclude that the field-like torque is the origin of the auto-oscillation. 



Although the oscillation in Fig. \ref{fig:fig2}(b) occurs around the $y$ axis, which is orthogonal to the easy ($x$) axis of the present system, 
the oscillation trajectory is similar to a small-amplitude oscillation observed in type-Y device \cite{liu12b}, 
where the magnetization oscillates around its easy axis parallel to the $y$ axis \cite{taniguchi17}. 
We should, however, note that the output power of the present system becomes larger than that from the type-Y device due to the following reason. 
In typical spin Hall devices, magnetic tunnel junctions (MTJs) are fabricated on the nonmagnetic electrode \cite{liu12b,fukami16}, 
where the ferromagnet in Fig. \ref{fig:fig1} is a free layer. 
The magnetization in the reference layer usually points to the direction parallel to the easy axis of the free layer to detect the magnetization switching. 
The shape magnetic anisotropy also prefers the easy axis of the reference layer parallel to that of the free layer. 
Therefore, the magnetization in the reference layer points to the $x$ direction in the present system, whereas it points to the $y$ direction in the type-Y device. 
The output power is proportional to the projection of the magnetization $\mathbf{m}$ to the direction of the magnetization in the reference layer. 
In the present system, $m_{x}$ varies as $-1 \lesssim m_{x} \lesssim 1$, whereas $m_{y}$ in typical type-Y devices varies as $0 \lesssim m_{y} \lesssim 1$ \cite{taniguchi17}. 
The difference comes from the fact that the magnetization in the present system oscillates around the $y$ axis, which is orthogonal to the in-plane easy axis, 
whereas the magnetization oscillation in type-Y devices occurs around the easy axis. 
As a result, the emission power of the present system could be larger than that from type-Y devices, 
although the magnetization oscillates around the in-plane axis. 


Figures \ref{fig:fig2}(d)-(f) show the time evolution of the magnetization, the dynamical trajectory in a steady state, and the Fourier transformation of $m_{x}$ 
for the current density of 160 MA/cm${}^{2}$. 
The spin-transfer torque due to a relatively large current moves the magnetization toward the $z$ direction and excites an out-of-plane auto-oscillation. 
The oscillation frequency shifts to higher frequency region and becomes 4.4 GHz. 
Since $m_{x}$ also varies as $-1 \lesssim m_{x} \lesssim 1$ here, a large output power could be expected. 
We also note that an out-of-plane oscillation as such does not occur in type-Y \cite{liu12b} 
because the field-like torque in this geometry prefers the in-plane oscillation. 



\begin{figure}
\centerline{\includegraphics[width=1.0\columnwidth]{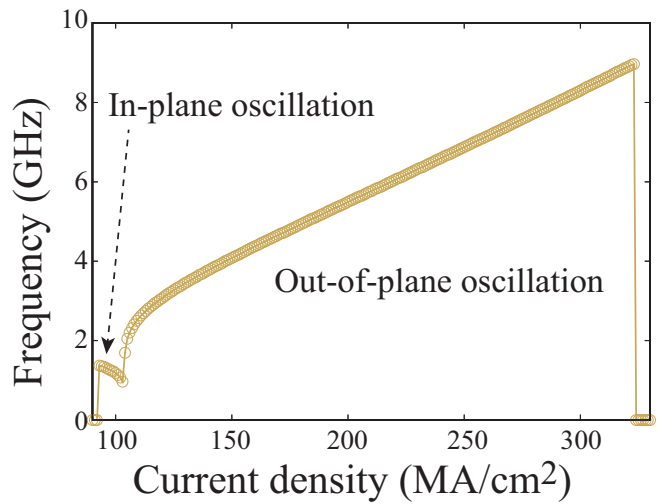}}
\caption{
         Dependence of the frequency of the magnetization oscillation on the current density. 
         The field-like torque is $\beta=0.3$. 
         The small and large current regions correspond to the in-plane and out-of-plane oscillations shown in Figs. \ref{fig:fig2}(b) and \ref{fig:fig2}(e), respectively. 
         \vspace{-3ex}}
\label{fig:fig3}
\end{figure}




Figure \ref{fig:fig3} summarizes the frequency corresponding to the first main peak in the Fourier space. 
It is zero when the magnetization stays near the initial state ($j \le$ 92 MA/cm${}^{2}$) or saturates to the $y$ direction ($j \ge$ 324 MA/cm${}^{2}$). 
The frequency becomes finite when the magnetization shows an auto-oscillation. 
In the small current region $93$ MA/cm${}^{2}$$\le j \le$ 103 MA/cm${}^{2}$, the frequency decreases with increasing the current, 
where the magnetization oscillates around the $y$ axis, as shown in Fig. \ref{fig:fig2}(b). 
On the other hand, the frequency increases with increasing the current 
in the large current region 104 MA/cm${}^{2}$ $\le j \le$ 323 MA/cm${}^{2}$, where the magnetization oscillates around the $z$ axis, as shown in Fig. \ref{fig:fig2}(e). 
The maximum frequency reaches close to 10 GHz. 
We emphasize that no external magnetic field is used in this study, 
whereas the previous works on the auto-oscillation in spintronics devices have frequently used external magnetic field \cite{kiselev03,rippard04,sankey05,krivorotov07,houssameddine07,kubota13}. 
We also note that the relation between current and frequency for different values of $\beta$ is shown in supplementary material. 
The supplementary material also provides an analytical formula of the current density to excite an in-plane oscillation. 



\begin{figure}
\centerline{\includegraphics[width=1.0\columnwidth]{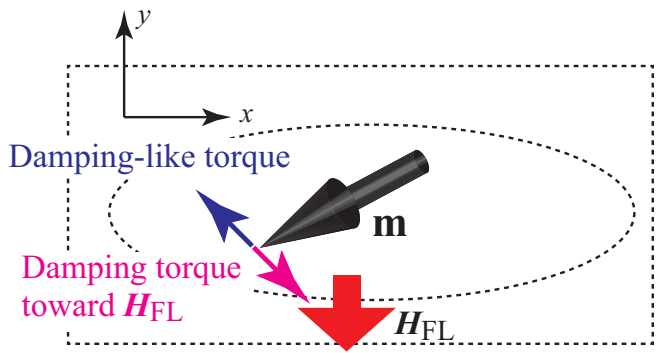}}
\caption{
        Schematic illustration of top view of the system. 
        The red arrow indicates the direction of an effective field $\mathbf{H}_{\rm FL}$ associated with the field like torque. 
        The pink arrow indicates the direction of the damping torque pointing in the direction of $\mathbf{H}_{\rm FL}$, 
        whereas the blue arrow indicates the direction of the damping-like torque. 
        The dotted lines represent the shapes of the top ferromagnet and the bottom nonmagnet. 
         \vspace{-3ex}}
\label{fig:fig4}
\end{figure}



We notice that the auto-oscillation is excited when $\beta$ is positive. 
In fact, auto-oscillations have not been observed in this geometry when the field-like torque is absent \cite{fukami16,taniguchi19,taniguchi20} or $\beta$ is negative \cite{shiokawa21}. 
This argument is also confirmed from the critical-current formula to excite the auto-oscillation shown in supplementary material, 
where the critical current is inversely proportional to $\sqrt{\beta}$, and therefore, becomes finite only for a finite $\beta$. 
Let us explain its physical meaning here.
For convention, we assume that $j>0$ and $\vartheta<0$, as used in the above calculations, 
although the following discussion is satisfied even when the sign of $j$ and/or $\vartheta$ are reversed. 
Auto-oscillations occurs as a result of the balance between the spin-transfer torque and the damping torque \cite{bertotti09}, 
where the spin-transfer torque refers the damping-like torque given by $-[\gamma\hbar\vartheta j/(2eMd)]\mathbf{m}\times(\mathbf{e}_{y}\times\mathbf{m})$ in Eq. (\ref{eq:LLG}). 
The damping torque induces a relaxation of the magnetization to the direction of the effective magnetic field. 
We should remind that the field-like torque provides an effective field pointing in the negative $y$ direction because $\beta$ is positive; 
see Fig. \ref{fig:fig4} where $\mathbf{H}_{\rm FL}=[\hbar \beta\vartheta j/(2eMd)] \mathbf{e}_{y}$ extracted from Eq. (\ref{eq:LLG}) represent the effective field associated by the field-like torque. 
Therefore, the damping torque acting on the magnetization $\mathbf{m}$ has a projection to the negative $y$ direction, as schematically shown in Fig. \ref{fig:fig4}. 
On the other hand, the spin-transfer torque moves the magnetization to the positive $y$ direction, as shown in Fig. \ref{fig:fig4}. 
Accordingly, there is a balance between the spin-transfer torque and damping torque, resulting in an excitation of sustainable auto-oscillation. 
If $\beta$ is negative, however, both the effective field associated by the field-like torque and the spin-transfer torque points to the positive $y$ direction. 
In this case, these torques do not cancel each other, and thus, auto-oscillation cannot be excited. 
Therefore, auto-oscillation can be excited only when $\beta$ is positive. 


In summary, we showed that an out-of-plane auto-oscillation of the magnetization can be excited in spin Hall devices with an in-plane magnetized ferromagnet. 
The origin of the auto-oscillation is the field-like torque with an appropriate sign. 
The field-like torque provides an effective field pointing in the direction orthogonal to the in-plane easy axis. 
The damping torque toward this effective field balances with the damping-like torque and excites sustainable oscillation. 
The oscillation found here has the following advantages for practical applications. 
First, a large amplitude oscillation results in a large emission power. 
Since $m_{x}$ varies in the range of $-1 \lesssim m_{x} \lesssim 1$, the emission power could be maximized in the present geometry. 
Second, no external magnetic field is necessary. 
This is because the field-like torque provides an effective field to sustain the oscillations. 
Third, the oscillation is excited in a ferromagnet with the easy axis parallel to the current. 
The type-X structure has a narrower injected-current cross-section comoared with type-Y devices;
thus, it can be used to reduce power consumption of application applications such as non-volatile memory \cite{fukami16}. 


See supplementary material where the magnetization dynamics at finite temperature, 
the current-frequency relation for a variety of $\beta$, and the analytical formula of the threshold current density to excite the in-plane oscillation are summarized. 


The author acknowledges to Masamitsu Hayashi, Yohei Shiokawa, Shinji Isogami, Takehiko Yorozu, Tomoyuki Sasaki, and Seiji Mitani for valuable discussion. 
This work was supported by funding from TDK Corporation. 


\section*{Data availability}
The data that support the findings of this study are available from the corresponding authors upon reasonable request.

\end{document}